\documentclass{LT23auth}

\usepackage{graphicx}

\begin{document}

\begin{frontmatter}


\title{Explanation of the tunneling phenomena between the edges of
two lateral quantum Hall systems}

\author[address1,address2]{Shinji Nonoyama},
\author[address1]{George Kirczenow}

\address[address1]{Department of Physics, Simon Fraser
University, Burnaby, B.C., Canada
V5A 1S6}

\address[address2]{Faculty of Education, Yamagata
University, Yamagata 990-8560, Japan}


\begin{abstract}
We identify the physics behind the results of recent  measurements 
[W. Kang {\em et al.}, Nature {\bf 403},
59 (2000)] of electron
transfer between the edges of two two-dimensional electron systems (2DES). We
find that a consistent explanation of {\em all} of the observed 
phenomena is possible
if the barrier between the 2DES is surrounded by a strong potential well
that supports quantum railroads of edge channels that, in the presence of
disorder, exhibit directed localization.
\end{abstract}

\begin{keyword}
tunneling; quantum Hall system; disorder
\end{keyword}
\end{frontmatter}

\begin{figure}[btp]
\begin{center}\leavevmode
\includegraphics[width=0.9\linewidth]{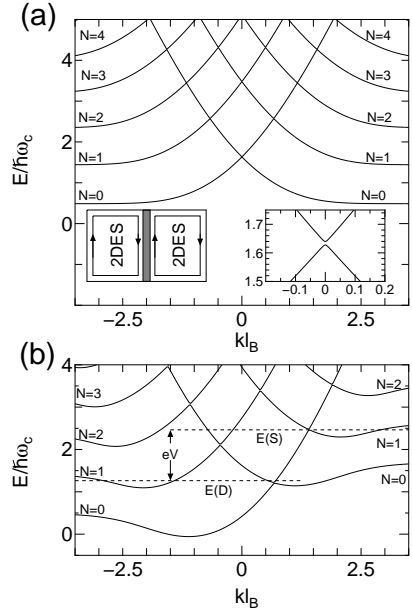}
\caption{
Landau levels $N$ = 0,1,2,...
become edge channels near barrier. a, Edge state energies
$E$ at zero bias vs. electron wave vector $k$ along barrier
in the simplest model of 2DES's and barrier.
Left inset: 2DEG's, barrier and edge
channels. Right inset: Energy gap at edge channel crossing.
b, Edge channel energies for bias $V$ between the
2DES's. Dashed lines are source and drain Fermi energies.
}\label{disp}\end{center}\end{figure}

Measurements of quantum tunneling through a barrier between the edges
of two two-dimensional electron systems (2DES) in a
transverse  magnetic field (see the left inset
of Fig.~\ref{disp}(a)) have revealed a
richness of puzzling phenomena~\cite{Kang}.
The energetics of the simplest model of this system that omits
disorder and electron-electron interactions is shown in Fig.~\ref{disp}(a).
In this model, for a long, high barrier electrons having energies 
within the gaps that occur at
Landau level crossings (see right inset of Fig.~\ref{disp} (a)) are transmitted
through the barrier while electrons at other energies are not~\cite{Kang}.
As was pointed out in Ref.~\cite{Kang} the predictions of
this simple model disagree with the data: The differential
conductance peaks at zero bias that are the experimental signature of tunneling
persist over ranges of magnetic field far larger than expected from the very
small widths of the energy gaps. Also these peaks are observed at magnetic
fields larger than predicted by factors of 2-4. More elaborate models 
of this system have
been studied~\cite{prev}. However this work assumed~\cite{prev} that 
the 2DES's are {\em fully} spin
polarized when the first zero-bias conductance peak (which occurs at 
Landau level fillings
$\nu > 1$) is observed. This requires an anomalously large enhancement of the
spin splitting of the 2DES. The absence of features due to spin in 
the data~\cite{Kang} is also
difficult to reconcile with this assumption. Theories to 
date~\cite{prev} have either not treated finite
bias voltages at all or yielded qualitative inconsistencies with the 
data~\cite{Kang} in that regime.  Here we
propose an explanation of the experiment~\cite{Kang} that is based on 
the physics of directed localization in
disordered 1D waveguides known as ``quantum railroads"~\cite{Barnes}. 
In our theory the spin splitting is
smaller than the Landau level broadening and thus can be neglected as 
a first approximation. Unlike the previous
theories~\cite{prev}, we explain {\em all} of the features of the 
data of Kang {\em et al.}~\cite{Kang}.

We carried out computer simulations of tunneling through the barrier 
using a recursive Green's function
technique~\cite{NO,NK}. We found that in the presence of disorder the tunneling
takes the form of a dense array of extremely narrow transmission 
resonances. But at the experimental
temperature~\cite{Kang} the individual resonances are not resolved and
we find transmission peaks with widths comparable to $\hbar 
\omega_c$, broad enough
to explain the observed persistence of the tunneling features and the 
absence of features due to spin in the
data~\cite{Kang}.  However the transmission  maxima are not shifted 
significantly by disorder from the energies at
which the crossings in Fig.~\ref{disp}  (a) occur. Thus disorder 
cannot by itself explain the observed
positions~\cite{Kang} of the tunneling  peaks. We therefore
carried out Hartree calculations of the Landau level crossing 
energies assuming a positive charge density
$\rho$ to be present in the barrier such as might be introduced by doping the
barrier in the plane where it was cleaved during fabrication of the sample.
This yielded a downward shift of the crossings sufficient to explain the
observed positions~\cite{Kang} of the zero bias conductance peaks for 
$\rho=11\times10^{11} e$
cm$^{-2}$.

\begin{figure}[btp]
\begin{center}\leavevmode
\includegraphics[width=0.9\linewidth]{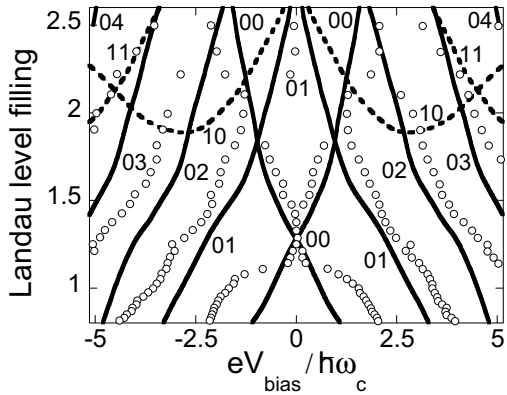}
\caption{
Comparison of theory (curves) and maxima
  of measured conductance\protect\cite{Kang} (circles)
for 2DEG's with density $2\times 10^{11}$cm$^{-2}$.
ij indicates tunneling from
(to) Landau level i(j).
}\label{bias}\end{center}\end{figure}

Edge channel energies calculated for this charged barrier
under applied bias are shown in
Fig.~\ref{disp}b. The energy of each edge channel now has
a minimum due to the electrostatic potential well near the barrier.
Therefore electrons can travel in opposite directions in edge
channels on the {\em same} side of the barrier and thus quantum railroads
exhibiting directed localization~\cite{Barnes} are formed.
The locations of the conductance maxima predicted by our
calculations of the edge channel crossings are compared with
the observed positions~\cite{Kang} of the conductance peaks
in Fig.~\ref{bias}. Dashed (solid) curves indicate edge channel
crossings at which tunneling is (not) expected to be
suppressed by competition between tunneling and the interchannel- and 
back-scattering of electrons
associated with directed localization in the quantum 
railroad~\cite{NK}. There is
an obvious one-to-one correspondence between the solid curves
and the loci of experimental conductance maxima and good
quantitative agreement at Landau level fillings $>$1.14 for
positive and small negative bias. However the solid curves do
not follow the bell-shaped structure at lower Landau level
fillings or exhibit the asymmetry seen experimentally
between positive and negative bias. We explain these deviations as 
manifestations of
the breakdown of the quantum Hall effect (QHE): This occurs at zero 
bias in 2DES's with {\em weak} spin polarization
(as in our theory) when $\nu$ falls below a value somewhat larger than 1.
When the QHE breaks down the 2DES becomes resistive and the measured 
bias voltage acquires a contribution
from the bulk of the 2DES {\em in addition} to that due to the 
barrier that is our theoretical bias voltage. Thus
when the QHE breaks down the experimental bias voltage of a 
conductance maximum begins to exceed the
theoretical one as happens abruptly below $\nu=1.14$ in Fig.~\ref{bias}.  This
explanation of the bell structure can be tested {\em directly} by measuring the
resistance of the 2DES. At high bias the QHE breaks down in a {\em 
sample-dependent} way at {\em all}
values of
$\nu$. This explains why at high bias the data~\cite{Kang} is 
asymmetric and why for high bias the
experimental maxima occur at somewhat higher bias values than the
solid curves in Fig.~\ref{bias}.

We thank W. Kang for helpful correspondence. This work was supported by the
Yamada Foundation (S.N.) and by the CIAR and NSERC (G.K.)

%
%

\end{document}